# The Impact of Incentives in the Telecommunications Act of 1996 on Corporate Strategies

By Martha A. García-Murillo

and

Ian MacInnes

*School of Information Studies,*

*Syracuse University CST 4-206*

*Syracuse, NY 13244-4100*

*e-mail: mgarciam@syr.edu*

*imacinne@syr.edu*

*Tel.: (315) 443-1829*

*Fax: (315) 443-5806*

# The impact of incentives in the Telecommunications Act of 1996

# on the business behavior of carriers

**Abstract:**

Rules are necessary to provide or shape the incentives of individuals and organizations. This is particularly true when free markets lead to undesirable outcomes. The Telecommunications Act of 1996 attempted to create incentives to foster competition. Ambiguity as well as the timing of the Act has led to delays in the clarification of rules and the rapid obsolescence of the document. The paper presents the strategies that common carriers adopted to try to tilt regulation in their favor, slow the entry of competitors, maintain their market leadership, and expand into other segments. Some of the strategies analyzed include lobbying efforts, court challenges, and lack of cooperation with new entrants.

**I. Introduction**

There are two common and opposite views of developments in the telecommunications industry after the passage of the Telecommunications Act of 1996. Some believe that competition is "developing at a robust pace"[1] while others are calling developments in the telecommunications market a "remonopolization" of the industry.[2] How can there be such deep differences of opinion on the same subject? The problem is that even five years after the Act passed, there is not a clear picture about the new structure of the industry or the amount of competition for each of the regulated segments. There are several reasons why the market has not yet settled, of which most are related to the incentives that the Act created and the actions that carriers took to maintain their market leadership positions.

---

[1] Steve Pociask, 'We've Come A Long Way ...; But Regulatory Barriers Have Placed The Future On Hold', *St. Louis Post-Dispatch,* December 27, 2000, Wednesday, B7.
[2] James Glassman, 'For whom Bells toll; Death of telecom competition', *The Washington Times,* December 27, 2000, A17.

To understand the impact of the Telecommunications Act it is not enough to evaluate developments since 1996. It is also necessary to examine the issues before then because it is not possible to determine the impact of a particular event without understanding the antecedents and the state of regulation before legislation was finally approved. Regulation of many of the issues that are addressed in the Act had already been part of the Federal Communications Commission Agenda in prior years. An assessment of the Act therefore has to be put into that context to determine if the problems and uncertainty that the industry was facing were actually resolved by the law. This paper analyzes events in the telecommunications industry from 1991 to 2001. The objective is twofold: (1) to provide a systematic analysis of events by coding and classifying information that could help in understanding the impact of the act; (2) a qualitative analysis of those events to understand the incentive system of the law as well as the process itself. This paper presents some of the factors that shaped the regulation and the strategic behavior that these rules created once they were put in place.

### II. Incentives

Incentives have been the realm of psychologists and economists. Psychologists have focused primarily on human behavior while economists have expanded their models to include firm behavior as well. In the telecommunications arena, economic analysis of the law focuses primarily on the economic incentives embedded in the economic models but has paid little attention to the incentive systems concerning competitive strategy, which is the realm of senior management.[3] In the business literature, incentives are an important element of this field's scholarly work. The focus nonetheless is on the effect of a company's monetary incentives on its employees' behavior. Compensation schemes are an example of this. Strategic corporate behavior is more closely related to the type of analysis that this paper presents although studies in the field of strategic behavior have not emphasized incentives and regulation. The contribution of this paper is in making an initial attempt at understanding such relationships.

---

[3] An excellent example of this type of research is Jean Tirole and Jean-Jacques Laffont, *Competition in Telecommunications. First ed, Munich Lectures in Economics,* Cambridge, MA: MIT Press, 2000.

Theories in cognitive psychology that can be extrapolated to explain firm behavior are expectancy value theory, which explains behavior [B] as a "function of the expectancies [E] one has and the value [V] of the goal toward which one is working"[4]

$$B = f(E \times V)$$

This means that people with multiple options will tend to choose the one that has the greatest value of the combination of expected success and value. This theory is relevant to businesses who, constrained by their budgets, can only pursue a few strategies.

Psychologists also make a distinction between internal and external incentives. An internal incentive occurs when the source of the motivation to do something comes from within the individual. An external incentive, on the other hand, occurs when the motivation comes from or is generated by a party aside of the subject. This distinction is useful within the context of regulated industries because many of their behaviors will be affected by external factors such as regulation. Internal versus external incentives serves as a framework in part V of this paper (see Tables 1 and 2).

Only recently have regulations focused on incentives. Normally regulation relies on what Barry Mitnick[5] calls regulation by directive and what theories of regulation have termed command and control. This type of regulation is simple to the extent that legislators need only to set in place requirements to which the regulated have to comply. Without an accurate understanding of industry and firm circumstances, prescribed solutions may lead to serious inefficiencies. Requirements on companies, "in addition to their cost-ineffectiveness, command-and-control approaches discourage technological innovation. When regulators effectively demand the use of a particular type of scrubber or other technology, industry will generally have no incentive to develop new or better control technologies."[6]

---

[4] "Incentive motivation," in *Encyclopedia Britannica On-line,* http://www.search.eb.com/bol/search?type=topic&query=incentive&DBase=Articles&x=17&y=3, September 21, 2001.
[5] Barry M. Mitnick, *The Political Economy of Regulation*, First ed. (New York: Columbia University Press, 1980), 342.

Unlike command and control regulation, incentive-based regulation attempts to modify behavior by changing the reward system, which would normally result in compliance with little or no monitoring from the regulator. Because of this, incentives have attracted the attention of regulators who, as a result, have written rules that try to incorporate them. Under command and control regulation, rules are directed as instructions or mandate particular conduct. Under incentive regulation a regulated firm has a set of alternatives of which some are more attractive than others. Also under incentives the regulated decides the course of action that best fit its circumstances, which ideally should also accomplish what the regulator wants. Wilkins and Hunt present an example that helps make this distinction clearer. As they describe, if Congress wanted to reduce the amount of air pollution it could use a command and control approach where it could dictate the type of smokestack that must be used or a standardized method for all of them. This type of regulation does not take into consideration differences across companies. While some of them will not need to make many adjustments to fulfill the standard, others will need to make substantial investments, perhaps to replace older equipment. An incentive system would give companies greater flexibility.

This paper presents some information about the Act itself but more importantly the incentives that were created from the process of developing rules for this industry. Unlike other studies in telecommunications regulation, this paper does not evaluate the incentives embedded in the economic models of the law but instead the strategic behavior of firms during the process of regulatory change.

**III. Methodology**

This research analyzed new events that were subsequently coded to gather more generalized data about companies and issues related to the Telecommunications Act.

---

[6] Timothy A. Wilkins and Terrell E. Hunt, "Agency Discretion and Advances in Regulatory Theory: Flexible Agency Approaches Toward the Regulated Community as a Model for the Congress-Agency Relationship," *George Washington Law Review,* 63, April, 1995, 485.

**Sample selection.** Two different coding schemes and news events were used for two of the sections in this paper. The first coding included the analysis of news events from two Regional (Bell) Holding Companies (RHCs). The reason why RHCs were selected was because they are more heavily regulated than their long distance counterparts. A random selection of news events was selected from the 1996-2001 period. In this case only the after Act period was selected simply to determine the impact that regulation has had on the activities of these companies. Because the focus is exclusively on the Act, the authors did not consider: news pertaining to the activities of these companies in other countries; activities regarding affiliates such as Lucent; society related activities such as funding a school event; and contract awards and personnel appointments. None of the aforementioned exclusions appear to be related to the carriers' market strategies. Although it may be argued that news are outlets that companies used to develop positive images, it is difficult to determine what is manipulation and what is real. On average a large enough amount of news would cancel out bias related to manipulation. The sample size was determined using a formula from Hinkle, Wiersma and Jurs.[7]

For sections VI and VII of this paper the selection of news was only of events related to interconnection and universal service. These were the only two issues analyzed in this paper because there were considered the most important under the many others for which the FCC needed to established rules.

**Coding criteria I**. The codification of these events consisted first of "1s" and "0s" to determine if the source of the action was due to an internal (coded as 1) decision or in response to an external requirement (coded as 0). An external incentive was selected if the company was forced to do something by an outside agent. For example, a regulatory body could force a carrier to improve its service by making it subject to fines. Internal incentives pertain to a company's own initiative to improve its market share, revenues, customer service, or other factors. The second type of code addresses the reason for such a decision. Six reasons were identified: (1) a regulatory requirement, (2) a regulatory requirement with negative consequences for the industry, (3) internal corporate initiative, (4) corporate initiative with negative consequences for consumers, (5) newly available technology, (6) a response to a competitor's action.

---

[7] William Wiersma, Dennis E. Hinkle, and Stephen G. Jurs, *Applied Statistics for the Behavioral Sciences*, First ed. (Boston: Houghton Mifflin Company, 1998).

In the coding process if it was not clear what the source of such action was, a series of options were outlined and then the one that was the most likely was selected. For example, a carrier could begin offering Internet access to a certain community. There are at least two reasons why they may have decided to do this. First, the *technology* could be more prevalent and they could offer it. Second, *competition* could force them to offer this service. The code selection would be as follows: assume, for example, that the technology was available but there was no competition. In this case the company could have simply not entered the market and be content simply continuing with the services that it is currently providing. Although this scenario is possible, providing Internet access is a service that, regardless of the existence of competition, they would like to provide because of the potential added revenues that can be generated. Assuming now that the company could have added this service but the technology did not exist then it could not have offered it even if it so desired. Under these circumstances the code was attributed to technology as a catalyst for service expansion for the carriers. It the technology was developed before 1996 then the source of the decision was attributed to the corporation. Rate decreases were coded as corporate because it is not clear if the reduction is a result of competition or of tariff reorganization within the company. If the company did not develop the technology, the event was coded as a technology source because it could not have provided such a service if they had not had it available. If the technology was developed inside the company then it was considered a corporate source. Activities that carriers were not allowed to provide under previous regulation but that they are now able to offer were attributed to *regulation.*

**Coding criteria II.** The criteria for sections VI and VII of this paper are presented in appendix one and two of this paper. When the news item only made a tangential comment about the issue of concerned, it was not coded.

**Reliability.** Intercoder reliability was analyzed by drawing a random subsample of news events from the overall sample, based on a formula by Lacy and Riffe.[8] The two authors coded the sub-sample. The percentage agreement and as the kappa value are indicated in each of the tables that were produced for this research.

**Selection of sources.** To reduce the possibility of bias due to the variability of sources from databases that combine publications, the authors chose to identify exclusively publications that focus on telecommunications alone. Nine sources that focus on this industry alone were identified. From this original list we selected those that were published for the entire period under study. There were only two periodicals that were published during the 1991-2001 period. One was *Informa Telecoms,* which had to be excluded because it focuses primarily on international news. The remaining source was *Communications Daily*.

**Selection of search terms.** To reduce the possibility of bias due to search terms the authors conducted sample searches to find all possible terms used in the industry press to refer to the same issue. Additionally the terms used in the Telecommunications Act were also selected as search terms for the particular issue under study. Once a complete set of search terms was identified, these were combined as the search term to be used in the database.

**Selection of issues for the coding of parts VI and VII.** Only issues that were unsettled by the Act were selected, specifically interconnection and universal service. Unsettled refers to the issues that the Act directed the Commission to define the rules. The reason for focusing on these issues alone is because without clear rules in the Act companies still have an opportunity to influence the FCC to develop rules that favor them. It is assumed here that in the absence of specified rules, carriers will have an incentive to engage in activities that will influence the rules in their favor.

**IV. The Telecommunications Act of 1996**

---

[8] Stephen Lacy and Daniel Riffe, "Sampling Error and Selecting Intercoder Reliability Samples for Normal Content Categories," *Journalism and Mass Communication Quarterly,* Vol. 73, no. 4 (1997): 963-973.

The Telecommunications Act is a complex document that attempts to accomplish multiple things. Its primary goal is deregulation of the telecommunications, cable, and broadcasting industries. This process is complicated because it encompassing all of the industries under the umbrella of the FCC. While the Act mentions many issues it does not provide specific rules for many of them. Given such limitations, the Act includes instructions to the FCC for the implementation of rules on issues that could only be outlined in the 1996 Act. Like many laws the Act contains clauses that are aimed to prevent regulators from overstepping their powers against the regulated. There are thus a number of clauses that indicates the things that government agencies can or cannot do. Similarly, as there are clauses that limit agencies demands to companies, there are also clauses intended to extend their privileges. While the first one is intended to prevent regulatory abuses, the second is intended to provide a mechanism for regulators to issue additional rules in case the existing rules fail to address a problem.

The existence of clauses intended to extend the rights of regulators could pose problems in the future because companies, knowing the discretion of these government agencies, will have an incentive to lobby to have rules written in their favor. As can be seen in Table 3, after the Act there were multiple challenges to orders that were being developed as well as to those that had already been issued. Additionally there are clauses where the Commission is given discretionary powers to extent the content of the law but at the same time give State commissions similar powers. While the intent of these clauses is to preserve the federalist spirit of the law, it leads to multiple problems.

Having states outline their own rules can give carriers multiple forums to play commissions against each other so that regulators outline rules that most benefit them. Alternatively, assuming that they are unable to get what they want they may simply operate and leave their most innovative initiatives to those states that have less stringent rules, effectively forum shopping. Differing rules between the federal and state commissions could also lead companies to comply with multiple requirements, effectively increasing their operating costs and indirectly leading to higher rates for their services.

Even though the National Association of Regulatory Utility Commissions provides an outlet for state commissions and federal regulators to coordinate their work, there have been cases in the course of the implementation of the Act where the differences between the two levels of government could not be reconciled, leading to lengthy court cases. This was the case of the interconnection rules, which is presented later in the paper. The Act also includes clauses designed to protect the regulated. Such clauses specify things that companies can do, may want to do, and should not be forced to do protecting them from regulators' unreasonable requests. Like any other rule there are also administrative elements that determine procedures to follow.

Among the issues that Congress left undefined were Sec. 227c, Protection of subscriber's privacy rights; Sec. 227d, Technical and procedural standards for facsimile, artificial or prerecorded messages; Sec. 228, Regulation of pay per call services and the possible extension of rules to data per call; Sec. 229, Communications assistance for law enforcement; Sec 245, Universal Service rules; Sec. 251d, Interconnection implementation rules; Sec. 257, Elimination of market barriers; Sec. 259, Infrastructure sharing regulations; and Sec. 276, Provisions for payphone services. These sections correspond only to Title II Common Carriers. Many of the issues that needed to be defined were highly complex and controversial, which must have left the staff at the Commission thinly spread. It is not surprising that some of these issues took months and even years to get settled. This paper focuses on the key issues of interconnection and universal service. These issues are the most important because they have the most direct effect on competition in the industry.

Negotiations for legislation like the Telecommunications Act can be lengthy and complicated because of the diversity of issues and the strong influence and power that companies in the affected industries had. Great contention existed between the RHCs and the long distance carriers about when and how each would be permitted to enter the other's market.[9] Companies organized themselves to achieve greater lobbying strength. The RHCs formed the Alliance for Competitive Communications and complained about the development of new regulation rather than the deregulation that had been promised.

---

[9] "Republicans May Have Sunk Telecom Bill For This Year; Congress: But A Panel Is Already Working To Gain The Support Needed To Pass It As Soon As Possible," *Los Angeles Times*, Section: Business, December 23, 1995, D-1.

The long distance carriers formed the Competitive Long Distance Coalition and were pleased with the law.[10]

### V. An overview of the impact of the Act

The objective of this paper is to analyze the incentive system and the regulatory process before and after the passage of the Telecommunications Act of 1996 and the impact that it has had on company strategies. To accomplish this the research was conducted in several stages. First an analysis of news events about carriers' activities was coded to determine the extent to which regulation had an effect on their daily operations. This first analysis was primarily exploratory to set up a context for the Act. In particular the intent was to see if rules were developed to such an extent that it was affecting the corporate behavior of carriers. The tables below present the results of this exploratory analysis.

Table 1

**Bell South Analysis of Decisions**

| | | Source of decision | | | | | | |
|---|---|---|---|---|---|---|---|---|
| | | Regulation | Regulation negative | Corporate decision | Corporate negative | Technology | Competition | Total |
| External incentive | Count | 14 | | | | | 1 | 15 |
| | % of Total | 10.3% | | | | | .7% | 11.0% |
| Internal incentive | Count | 21 | 4 | 25 | 9 | 10 | 51 | 121 |
| | % of Total | 15.4% | 2.9% | 18.4% | 6.6% | 7.4% | 37.5% | 89.0% |
| Total | Count | 35 | 4 | 25 | 9 | 10 | 52 | 136 |
| | % of Total | 25.7% | 2.9% | 18.4% | 6.6% | 7.4% | 38.2% | 100.0% |

$\chi^2 = 55.3$ Sig. $= .000$

Table 2

**Ameritech Analysis of Decisions**

| | | Source of decision | | | | | | |
|---|---|---|---|---|---|---|---|---|
| | | Regulation | Regulation negative | Corporate decision | Corporate negative | Technology | Competition | Total |
| External incentive | Count | 29 | | | 1 | | 1 | 31 |
| | % of Total | 18.5% | | | .6% | | .6% | 19.7% |
| Internal incentive | Count | 28 | 4 | 61 | 17 | 3 | 13 | 126 |
| | % of Total | 17.8% | 2.5% | 38.9% | 10.8% | 1.9% | 8.3% | 80.3% |
| Total | Count | 57 | 4 | 61 | 18 | 3 | 14 | 157 |
| | % of Total | 36.3% | 2.5% | 38.9% | 11.5% | 1.9% | 8.9% | 100.0% |

$\chi^2 = 40.4$ Sig. $= .000$

---

[10] "No 'Date Certain'; Senate Democrats Release Alternative Telecommunications Reform Bill," *Communications Daily,* Vol. 15, No. 32, February 16, 1995, 1.

These tables show that the vast majority of decisions made by these companies were motivated internally. This means that corporate actions result from the initiatives of company executives and not from a regulatory agency mandating them to do something. Even though both of these are local carriers they reacted differently to their market circumstances. Bell South, for example, made decisions that were intended to address competitive issues while for Ameritech this was of little concern. On the contrary the later one was occupied instead with new corporate initiatives. Regulation had an impact on their activities although Ameritech seemed to be more affected with 36% of the news events analyzed attributed to regulation. The differences on the impact that regulation has on each of them vary depending on the states where they provide services. Some public utility commissions (PUCs) are more interventionist than others and this could therefore lead to some companies being more affected than others by regulation. The broad analysis of these events simply indicates that regulation indeed has constraining effects on companies' behavior. Since we made the distinction of negative effects, regulation appears to be primarily for the benefit of society in spite of the constraints that it poses on companies. There were relatively few events that were considered to have negative effects. These were 2.5% for Ameritech and almost 3% for Bell South.

The purpose of this analysis is to determine the degree to which regulation has an effect on the daily activities of these companies. Some people may argue that regulation should be implemented such that it is transparent and does not dominate the company's market decisions. When regulation leads companies to make most of their decisions comply with government rules, the danger is that inefficiencies in the industry will emerge that could threaten its long-term economic health. In the case of the telecommunications industry regulation has had a substantial as is shown in Tables 1 and 2.

**VI**. **Interconnection**

To understand corporate behavior before, during, and after the passage of the Act, it is important to be aware of the circumstances surrounding these regulatory developments. In the early 1990s computing technology was evolving rapidly and, with the emergence of the graphical Internet browser companies began to see additional revenue opportunities. Similarly at the beginning of the Clinton Administration, Vice-president Al Gore was a strong supporter of the development of an information superhighway. Part of this vision included the widespread ability to exchange computerized video images, sophisticated graphics, sound, high-speed data and video transfers, conferencing and video on demand. In the summer of 1992 Al Gore and George Brown introduced the Information Infrastructure Technology Act and in 1993 the Clinton Administration issued a report called "Information in the 21$^{st}$ Century."[11] The enthusiasm and support that the administration showed for technology, and specifically telecommunications, led to a heightened interest in investments related to advanced services. It was at this time that telecommunications companies started their video on demand trials and telecommunication companies became more interested in expanding their services beyond their own markets.

Provision of telecommunication services nonetheless was something that attracted the attention of other companies, in particular carriers that were serving private corporations. In a market that seemed full of opportunities, it was natural for these carriers to want to expand their services to residential customers as well. This nonetheless required the cooperation of incumbent carriers that would have to allow them to interconnect with their facilities. There were attempts by incoming carriers to obtain interconnection even before 1996. Teleport, for example, was one of the first companies to request interconnection with an RHC. This was so that it could expand services offered to its New York clients.[12] Companies such as Teleport pressured public officials to issue regulations that would make it easier for them to obtain access to the incumbent's infrastructure.

In the interest of fostering competition some states issued regulations to mandate interconnection. As early as 1991 the FCC had already issued orders to mandate interconnection. Table 3 shows that several interconnection agreements had already been negotiated among companies before the 1996 Act.

---

[11] "Prepared Statement Of John Sculley, Chairman, Apple Computer and Chairman, Computer Systems Policy Project Before the House Telecommunications and Finance Subcommittee," *Federal News Service,* January 19, 1993.
[12] Diane S. Boiler, "The Year in Review," *Public Utilities Fortnightly,* January 1 1991, 44.

At this time the RHCs wanted to offer new services and petitioned state PUCs for permission to enter the long distance market. In exchange they offered to allow access to their networks. At this time incumbent telecommunication carriers were heavily regulated and considerably restricted in the services that they were allowed to provide. Because of their monopoly status primarily at the state level it would have been difficult for them to simply request to be allowed into the long distance service, which was the first area where they wanted to expand. They therefore had an incentive to offer in exchange access to their networks to requesting communication carriers.

Most interconnection requests were handled at the state level and, as a result, several PUCs issued regulations to foster competition, sometimes even at the local level as was the case in Chicago which proposed in 1992 a "Telecommunications Free Trade Zone,"[13] an initiative that was first suggested by Ameritech. At that time 17 states already allowed competition in local exchange service while 18 were formally considering allowing competition and four allowed it informally.[14] While there were some states that issued regulations others refused to address the issue and requested instead that the FCC resolve the matter.

Although there may have been some companies that, under the possibility of being allowed to expand their services and markets, may have an incentive to open their networks, there are also signs of refusal on the part of the RHCs that made it difficult for entrants to gain access to national networks. At that time the excuse from incumbents was that it would cost them too much to improve their networks.

Because rules were not yet developed, companies had to present their plans to either state commissions and on occasions to the FCC. Decisions then had to be made with respect to their requests regarding provision of enhanced services or even entering the long distance market.

---

[13] Communications Daily, February 4 1992, 4.
[14] "Common Carrier Competition," *FCC, Common Carrier Bureau,* Spring 1995, 6, http://www.fcc.gov/Bureaus/Common_Carrier/Reports/FCC-State_Link/fcc-link.html

When the issue was brought to the FCC for consideration, the commission was already working on issues of interconnection concerning number portability for 800 numbers.[15] From a strategic perspective, 800 number portability was of interest to carriers because it would help them to keep customers that needed to maintain the same telephone number even after they moved. Additionally these were, for the most part, business clients with larger accounts than their residential counterparts. Interconnection with other companies was part of the concessions they had to make to satisfy a customer need. Maintaining these accounts was therefore a strong incentive for them to cooperate. RHC willingness to allow interconnection to would-be competitors turned into resistance. In the case of number portability, companies were able to maintain revenues but interconnection with competitors represented instead revenue decline. They therefore put greater emphasis on the costs of allowing such network connections. They also argued that that it would have negative effects in terms of network quality, finances, and job losses.[16] They emphasized that interconnection could threaten networks due to technical difficulties. Table 3 presents data about the number of interconnections denials that happened before and after the Act was passed.

Incumbent LECs lose the most when they allow interconnection. They receive lower revenues because of the expense of accommodating the technical needs of other carriers and then having to share the market. When one carrier opens its network while the others remain closed, the carrier that opened its market can lose considerable revenues. This is because of added competition in its home market without the ability to participate in the other carriers' closed networks. In the meantime the carrier that kept its market closed maintains its monopoly while extending its market. If they all decide to keep their markets closed they then remain as the dominant carriers with no challenges. The dominant strategy is for carriers to maintain closed markets

While the FCC was being pressured from competitive carriers to mandate interconnection, Congress was instead debating whether or not they should remove MFJ restrictions so that RHCs could enter the long distance market. In May of 1992 The

---

[15] "Callers Won't Notice Change; Phone Industry Drafts Transition To 800 Portability," *Communications Daily,* December 19 1991, 2.

[16] "Vote Seen As 3-1/2 to 1-1/2; Physical Interconnection Issue Creating Conflicts," *Communications Daily,* October 7 1992, 4.

House Commercial Law Subcommittee passed the Brooks MFJ Bill (HR-5096) that was intended to slow down the entry of RHCs to information services, manufacturing and long distance.[17]

The FCC was more open to the idea of deregulation than Congress In 1992 the FCC issued an order requiring "LECs with revenues of over $100 million annually to offer expanded interconnection to all interested parties, permitting competitors and high volume users to terminate their own special access transmission facilities at LEC central offices."[18]

Table 3

**Interconnection Issues**

| | | ISSUES | | | | | | | | Total |
|---|---|---|---|---|---|---|---|---|---|---|
| | | Interconnection agreement | Federal rules established | Court challenge | Organizational challenge | State rules established | Company refusal for interconnection | Individual company review | Federal rules to be determined | |
| Before the 1996 Act | Count | 25 | 21 | 5 | 9 | 11 | 11 | 28 | 68 | 178 |
| | % of Total | 6.3% | 5.3% | 1.3% | 2.3% | 2.8% | 2.8% | 7.1% | 17.2% | 44.9% |
| After the 1996 Act | Count | 50 | 8 | 25 | 38 | 12 | 40 | 29 | 16 | 218 |
| | % of Total | 12.6% | 2.0% | 6.3% | 9.6% | 3.0% | 10.1% | 7.3% | 4.0% | 55.1% |
| Total | Count | 75 | 29 | 30 | 47 | 23 | 51 | 57 | 84 | 396 |
| | % of Total | 18.9% | 7.3% | 7.6% | 11.9% | 5.8% | 12.9% | 14.4% | 21.2% | 100.0% |

$\chi2 = 91.02$ Sig. = .000

After the Act was passed things did not get easier but rather became more complicated. Prior to the Act state PUCs were handling some of the interconnections issues and were even able to intermediate some disputes. This nonetheless did not appear to be an ideal situation considering that without national rules, states had to devise their own criteria for interconnection. The Act therefore seemed like a solution to the emerging chaotic situation of state rules and carrier interconnection negotiations. While in theory the Act was supposed to settle the issue and provide guidelines for interconnection, it led instead to an even more chaotic situation. The 1996 Act left it to the FCC to determine the rules for interconnection, providing only general guidelines. Because these rules were not specified from the beginning there was ample room for interpretation. As can be seen in Table 3, many of the news events related to interconnection refer to the many aspects of interconnection that needed to be determined by the FCC. News events related to interconnection agreements doubled after the Act. This in fact meant that the interconnection negotiations that had been agreed upon before the Act were company-specific and did not address issues that had greater impact on the entire industry. Unlike these individual agreements, the Act had to be concerned with the greater impact of these agreements on competition, incentives for infrastructure investment, national technical and regulatory standards, and universal service.

---

[17] "Markey Gearing Up; Brooks Mfj Bill Passed, 10-6, By House Judiciary Unit," *Communications Daily,* May 29 1992, 1.
[18] *Communications Daily,* September 21 1992, 6.

Because the Act left the issue to be determined by the FCC, carriers had an incentive to lobby to interpret the law in their favor. Soon after the Act was passed and even before the Commission had an opportunity to draft these rules, the Association for Local Telecommunication Services (ALTS), which represents competitive access providers (CAPs), submitted a document entitled *Handbook for the FCC* where they specify their interpretation and the things that they expected the FCC to do.[19]

Even before the FCC issued its interconnection order there was considerable enthusiasm on the part both of the competitive local exchange carriers (CLECs) and the RHCs about the possibility of extending their markets. Bell Atlantic and Ameritech were among the first to submit their applications to both PUCs and the FCC to obtain approval to begin providing long distance service. This happened after only one competitor had signed an interconnection agreement with it. The company believed that this was enough to fulfill the interconnection requirements of Secs. 521 and 252. These were premature applications that were quickly denied. The Table 3 heading of individual company evaluation provides some indication of requests that were submitted to the FCC and state commissions regarding interconnection issues.

Once the order for interconnection was issued[20] there was considerable debate from both the CLECs as well as the RHCs. Initially it appeared as if the RHCs were willing to cooperate on interconnection agreements although they expected the authorization to provide long distance service to be almost automatic. Although it is not something that they explicitly recognized, it is possible that the frustration with the application process to offer interexchange access also contributed to their reluctance to facilitate interconnection to competing carriers. As the years passed there was an increasing number of court challenges. The issue became increasingly complex and confrontational. There were multiple issues where there were differences of opinion. The most important was the calculation of rates for unbundled network elements (UNEs). This was nonetheless only the beginning of a series of company and court challenges to the interconnection rules. It began with the formula used to calculate the rates and escalated to include issues such as property rights for third party equipment on collocation agreements, the quality of the networks to be leased, the network elements included in the UNEs list, the obligations and rights of wireless carriers, and the impact of leased networks on access charges and universal service. It was clear that Congress and the FCC underestimated the complexity of the issue. Table 3 presents some information regarding the number of news items that made allusion to court challenges before and after the Act. From a corporate strategy perspective one could argue that court challenges could in fact help to slow the entrance of a competitor. There were nonetheless issues that indeed had serious implications. Rates alone were crucial because a low tariff could have negative affect on their revenues and reduce the incentives for carriers to continue building a state of the art infrastructure and perhaps such low rates could have even threatened the survival of some of the incumbents. Low rates, as they argued could also eliminate incentives to foster facilities based competition. Other arguments, such as equating physical collocation with stealing property, were more questionable.

---

[19] *Communications Daily,* March 18 1996, 5.
[20] FCC, "The First Report & Order In the Matter of Implementation of the Local Competition Provisions in the Telecommunications Act of 1996," (http://www.fcc.gov/ccb/local_competition/fcc96325.html: 1996).

  Aside from rates, common carriers also tried to limit the network elements that they were obliged to lease. Originally the Commission prepared a list that included all network elements. Carriers naturally challenged this list in an effort to slow and limit the amount of competition that would otherwise come into their markets. The Supreme Court that reviewed the list requested that the Commission revise it to take into consideration the "necessary" standard and the "impair" standard of the Act.[21] Rather than revise parts of the list, the FCC instead made it longer by including conditioned loops as well as line sharing.[22] The list was challenged once more at the Eighth Circuit, where the court stated that the Act only required the unbundling of existing network elements and not of elements that may exist in the future. They also reasoned that incumbents are not obliged to "cater to every desire of every requesting carrier"[23] FCC officials at this time claimed that the Supreme Court had already decided on this issue and did not change their position. The dispute has continued.

---

[21] UNE Remand Order at 3745 ¶101.

Because the FCC established rules that tended to favor competitors over RHC incumbents one of the strategies that incumbent local exchange carriers (ILECs) utilized to prevent these rules from prevailing was to accuse the FCC of overreaching its authority by imposing rules at the national level when the Act had specified that states had the authority to devise their own interconnection rules.[24] The existence of clauses in the Act that gave both the FCC and state commissions power over regulation gave incumbents a reason to complain. Before the Act and interconnection orders were issued, the establishment of national rules was welcome. Once they were issued, states found themselves having to decide which rules to apply: their own or those issued by the FCC. This was the case in Florida, Ohio, and California, which had to put on hold their ruling processes to wait for the FCC to issue its interconnection order. Although the ILECs were the ones that first challenged the authority of the FCC to determine pricing rules for the entire country, state commissions followed up by making similar complaints. States that already had passed interconnection rules were concerned about their validity.

Other controversial issues included the formula used for interconnection rates and UNE deaveraging, which could lead to higher rates to subscribers in low density areas. Some commissioners were concerned that rural communities would face considerably higher rates than their urban counterparts if deaveraging was used. This was also unforeseen because the idea of deaveraging was simply to allow carriers to have lower rates for those areas where services provision was lower. This implied that some areas of the country would have lower rates for UNEs than others. Court challenges took such a long time to be resolved that PUCs had to resort to interim rates waiting for those cases to be sorted out.

---

[22] Line sharing refers to the unbundling of high frequency spectrum on a telephone line that is above the voice-band. This portion of the line is generally used to provide x-DSL services.
[23] Iowa Utilities Bd., 120 F3d 753, 819 n. 39 (8th Cir. 1997) aff'd in part and rev'd in part, AT&T v. Iowa Utils. Bd. 119 S. Ct. 721 (1999).
[24] See Sec. 251d3 Preservation of State Access Regulations.

In summary, incumbents did not expect to have their applications rejected and wanted to enter long distance markets quickly. Unable to enter long distance, they instead tried to limit the entry of competitors. Their strategies included first trying to influence the rules that would come out of the FCC and second challenging them. The court challenges cannot be seen as primarily frivolous as there were legitimate concerns. Incumbents were also successful at having PUCs support their cause. These legal battles effectively slowed the entrance of competitors for approximately four years until courts reinstated the authority of the FCC to determine the standards for interconnection. Because of this and all the contracts that were done on an interim basis, it is expected that more conflicts will arise at the time of renewal.

The process of interconnection has stabilized and has become more formalized having third party tests on incumbent infrastructure to determine if they are adequate for serving competitors. This prevents the negotiations from becoming too contentious and provides a less partisan evaluation of the incumbent's willingness to cooperate with competitors. These impartial evaluations serve them well when they submit their applications for the provision of long distance services.

Table 3 shows that before the Act was passed there were some interconnections negotiated but the signing of the Act opened the door to other interested parties that would not otherwise have requested entrance. It would have been difficult for small entrants to negotiate interconnection contracts without regulatory obligations imposed on incumbents. The regulation did not make it easier but there was at least some pressure to comply. The main problem is the incentive system that was embedded in the interconnection rules. It required incumbents to share an infrastructure that they built with would be competitors. This regulatory obligation contradicts rational business behavior and consequently there is very little interest in complying. While there were refusals for interconnection prior to the Act, there were also refusals afterward. The emergence of broadband networks has added yet another challenge to the FCC. Incumbents do not want this upgraded networks to be part of UNEs and asked for an exception. A more recent attempt to try to exclude broadband from interconnection agreements was from Bell South, which set up a separate unit for advance services alone that are not subject to regulation. By doing this they have effectively taken these services out of the UNEs. Interconnection of broadband networks is still to be decided.

### VII. Universal Service

Universal service concerns have been an issue of contention for the FCC since the early 1990s. At that time with the emergence of competitive carriers, public utility commissions became concerned about contributions to the universal service fund, which were only required of RHCs. Initial discussions were held regarding the services to be included under universal service support. Because universal service was related to interconnection, the telecommunications regulations issued by PUCs often addressed both issues. The close relationship between interconnection and universal service made both issues more complex. Rates for unbundled network elements would have to take into consideration contributions to the universal service fund if the FCC were to require them from all carriers including new entrants. Discussions that were taking place before the Act concerned the services that would or would not be included for support. While PUCs wanted to resist the support of emerging technologies as part of universal service, there was some interest on the part of carriers to include them. Telecommunications companies would have liked some of these services to be subsidized because this would increase their revenues. This has given them an incentive to lobby for their inclusion.

During discussions taking place at the state level, incumbent carriers argued that they would be at a disadvantage if they were required to contribute while new entrants were exempt and lobbied to prevent this.[25] Competitive carriers on the other hand argued that the interconnection charges that incumbents receive should be enough to cover the costs of providing services to rural areas.[26] LECs also argued that with fewer subscribers the resources of the universal service fund would be depleted. In spite of the complexity of the issue and strong lobbying from incumbents some states issued regulations on the subject. Table 4 presents information about the number of news events that made allusion to state rules established as well as the amount of news addressing company challenges to the law. This shows that several states had established rules even before the Act. For states without regulatory attempts companies often negotiated with commissioners to reduce regulations and in exchange offer to improve universal service provisions.[27]

---

[25] *Communications Daily,* September 27 1993, 3.
[26] *Communications Daily,* November 18 1994, 1.

The analysis of events shows that few rules were implemented at the federal level compared to the amount of activity that happened after the Act was passed. Carriers criticized the establishment of interim rules until a more comprehensive study was carried out. It is surprising to see in Table 4 that most of the news regarding regulation was about the things that still have to be done. This was true both before and after the Act. The few orders issued by the FCC addressed the increases in the contributions that carriers had to make to the fund. Similarly orders CC Doc. 91-141, Transport Phase I, and CC Doc. 80-286, which also addressed interconnection and competition issues, left universal service contributions untouched. Carriers that traditionally served rural communities complained that such rules opened the market to competitors and that their presence would lead to reduced revenues. They argued that this would negatively affect the provision of services to rural and some residential users.[28] From an incentives perspective it is clear that carriers would want to avoid losing the support that they had enjoyed up to that point. There was therefore considerable debate on the part of carriers providing services to rural communities to make modifications to the law such that they would not lose such resources after competition happened.[29] During the period prior to the Act some carriers objected to a model suggesting subsidies paid directly to carriers.

As can be seen by the $\chi^2$, the difference in the number of news items for the periods before and after the act is not significant. Although this is an overall test of independence it nonetheless gives some indication of the small differences that existed before and after the fact with respect to the areas of uncertainty. In this respect it could be said that the Act failed to resolve universal service issues.

---

[27] *Communications Daily,* May 18 1993, 4.
[28] "Most Large Carriers Affected; FCC Calls For Expanded Interconnection Opportunities In Switched Transport," *Communications Daily,* 1993, 1.
[29] *Communications Daily,* September 12 1994, 8.

Table 4

Universal Service Analysis of issues

| | | ISSUE | | | | | | | | Total |
|---|---|---|---|---|---|---|---|---|---|---|
| | | Federal regulation established | Court challenge | Organizational challenge | State regulation established | Contribution determination | Individual company evaluation | Federal rules to be determined | State regulations to be determined | |
| Before the 1996 Act | Count | 5 | | 11 | 3 | 3 | 8 | 39 | 12 | 81 |
| | % of Total | 2.2% | | 4.8% | 1.3% | 1.3% | 3.5% | 17.0% | 5.2% | 35.2% |
| After the 1996 Act | Count | 20 | 8 | 31 | 5 | 7 | 13 | 53 | 12 | 149 |
| | % of Total | 8.7% | 3.5% | 13.5% | 2.2% | 3.0% | 5.7% | 23.0% | 5.2% | 64.8% |
| Total | Count | 25 | 8 | 42 | 8 | 10 | 21 | 92 | 24 | 230 |
| | % of Total | 10.9% | 3.5% | 18.3% | 3.5% | 4.3% | 9.1% | 40.0% | 10.4% | 100.0% |

$\chi^2$ = 13.5 Sig. = .094 Kappa =

Universal service, like interconnection provisions, were left undetermined by the Act. As stated in Sec. 254:

> FEDERAL-STATE JOINT BOARD ON UNIVERSAL SERVICE.—Within one month after the date of enactment of the Telecommunications Act of 1996, the Commission shall institute and refer to a Federal-State Joint Board under section 410(c) a proceeding to recommend changes to any of its regulations in order to implement sections 214(e) and this section, including the definition of the services that are supported by Federal universal service support mechanisms and a specific timetable for completion of such recommendations.[30]

Even after the Act there were some state commissions that were still trying to develop rules for interconnection and universal service. California, for example, issued its universal service rules in August 1996.[31]

Because the rules for universal service still had to be determine by the FCC there were several reports sent to the FCC to influence the amount of contributions. GTE, Pacific Bell, and SNET were among the companies that prepared reports.[32] Similarly, hearings organized by the Commission were highly contested as particular carriers disagreed on the formula to be used to determine costs as well as the criteria used to determine which geographical areas needed subsidies.[33] Table 3 presents information about the number of news items that talked about the determination of contributions before and after the Act. It shows that these remain an issue of discussion after the Act.

---

[30] "Telecommunications Act of 1996," *Federal Communications Commission,* www.fcc.gov/telecom.html.
[31] *Communications Daily,* August 13 1996, 5.
[32] *Communications Daily,* May 20 1996, 5.; *Communications Daily,* June 17 1996, 3; and *Communications Daily,* July 5 1996, 5.
[33] "Industry Offers Views; Joint Board Contemplates Universal Service Approaches," *Communications Daily,* June 6 1996, 2.

Contention was also aroused among carriers because the incumbents argued that competitive carriers could already have been receiving a subsidy if they bought UNEs and are exempt from paying interconnection charges. The timing that was established to determine the rules for interconnection and universal service was also an issue of debate. In the Act rules for interconnection were due before those for universal service and carriers were concerned about the fund falling short of what was necessary to support US programs. In this case Bell South, which was advocating this view, wanted the Commission to maintain access charges until the rules for Universal Service were determined. This was obviously an argument to maintain such a revenue flow. Competitive carriers rejected this policy by saying that they would be paying double fees for UNEs and access charges. They instead argued that they would only pay to support the universal service fund and its two associated programs, Lifeline and Link-up.

In the discussion about allocation of funds a recommendation was made about granting funds to carriers based on pre-established criteria such as quality of service, level of interconnection, number portability, and the geographical extent of the network. This type of criteria nonetheless increases the monitoring costs of the regulator.

The Act was signed at a time when technology was changing rapidly and the Internet had become a major force in all types of social and commercial activities. This made it necessary to consider proposals to provide universal service funding to connect libraries, schools and rural health centers to the Internet. Similarly wireless technology was gaining support and regulators had to consider these carriers as possibilities for carriers of last resort (COLR) in rural communities.

In selecting carriers receiving support to serve rural communities, regulators encountered a dilemma. On the one hand they wanted to have competition but at the rural level they may have only one option. Having only one carrier meant that these communities would have no choice and the government would then be obliged to regulate its rates and services, something it wanted to eliminate with the Act. The calculation of costs to serve rural communities was also controversial. From an incentives perspective the ILECs would like their costs to be the ones used to determine the number of subsidies they receive to serve these communities. The long distance carriers on the other hand rejected this formula because they justifiably believed that the ILECs without much monitoring would have an incentive to inflate their costs. The long distance carriers therefore preferred proxy costs to be used to calculate these subsidies. Independently of the model used regulators were aware that all types of subsidy mechanisms had to be dealt with simultaneously to determine the amount of money that each of these carriers had to contribute. The process was more complex than expected and by June 1997 regulators were still looking at the models to be used for universal service and expected at the time that implementation would then take place by 1999.[34]

Interexchange carriers (IXCs) saw reform in universal service as an opportunity to eliminate access charges completely to be replaced instead with subscriber line charges. They wanted to have transfers made from carriers servicing high cost areas to those serving low cost ones.[35] In defense RHCs argued that access charges pay for universal service and that if they were to be modified they will have to be sufficient to cover these costs. Because of objection from IXCs, they propose to have an access charge that is not related to use but instead is a flat fee per line.

Added to the contention between the IXCs and the LECs, an additional set of players needed to be included in the discussion. These were information service providers, such as those offering Internet access, that were subject to less regulation. Because information service providers are not defined as telecommunications carriers they do not have to pay universal service fees. Some policymakers argued that the telephone companies would be subsidizing the computer industry if they received funds for Internet connections.[36]

Allocation of universal service funds was not something that affected carriers alone. Because part of the universal service plan was to wire schools, libraries, and health centers in rural communities, there was considerable disagreement between the states and the FCC. This is because each state allocates funds to schools differently and each state wanted to have the flexibility to do so in its own way. There was thus an incentive on the part of the state to try to keep the determination of subsidies to schools at the state rather than the federal level. There were nonetheless some CLEC organizations such as the American Communications Services Inc. (ACSI) that did not want states to be involved and submitted letter to the FCC asking for federal preemption of state rules.[37]

---

[34] "Federal-State Joint Board Members Critique Universal Service Order," *Communications Daily,* May 16 1997.
[35] "Might Encourage Gaming; Industry Praises, Pans GTE Auction Proposal," *Communications Daily,* August 7 1996, 2.
[36] "Senators Criticize FCC On Universal Service Support," *Communications Daily,* June 4 1997.

Another area where there are conflicting incentives is between small companies and large companies. While the smaller carriers would like to be completely exempt from regulation, larger carriers want them to have the same regulatory burden as anyone else. Although smaller carriers can be exempt, they nonetheless have to prove that they are small enough to be exempt.

Universal service issues were not adequately addressed in the 1996 Act. The lack of rules in the law itself provided carriers with opportunities to lobby for their interests and slow the process. A great incentive existed for them to challenge implementation regulations because of the possibility of losing universal service support and revenue from access fees. The confusion that existed before the Act continued to prevail afterward.

### VIII. Conclusion

The Telecommunications Act of 1996 was clearly necessary because it deregulated areas that would have remained stagnant if there had not been a loosening of the rules to allow carriers to enter other information service areas and eventually the interexchange market. The rules nonetheless did not make things easier as the law included segments that provided few guidelines. This provided ample room for carriers to engage in substantial lobbying to influence FCC regulations in their favor. Because the issues that Congress and subsequently the FCC had to deal with were highly complex, there was great debate. The implementation of the law led to even more confusion than there was before the Act passed. In this respect the law was not helpful. Technological evolution also contributed to making implementation difficult. As rules were being devised, new applications were developed that inevitably would have made any regulation obsolete at the time it was issued. On universal service, for example, while the commission was focusing primarily on support to rural communities, the rapid evolution of the Internet made it necessary to consider this infrastructure part of universal service concerns. Similarly, at the time the act was drafted broadband communications had not been implemented and were later an issue of contention when they had to be considered part of the UNEs list. The Act provided some relief with respect to interconnection but was inadequate on universal service issues.

---

[37] "ACSI Asks FCC To Preempt Authority Of Ark. PSC," *Communications Daily,* March 27 1997.

# Appendix 1

## Interconnection: operationalization of codes

| | |
|---|---|
| Interconnection agreement | An agreement is signed between two companies or with the help of a regulator. An incumbent carrier reports about its interconnection agreements. A carrier or PUC reports about interconnection agreements being made to date. |
| Federal rules established | The FCC issues an order related to the implementation of interconnection related issues. The FCC successfully clarifies an issue of contention. Congress passes a bill pertaining to interconnection. |
| Court challenge | Companies files a lawsuit against either the FCC or PUCs. |
| Organizational challenge to the FCC | When an individual company, an association representing carriers, or a consumer group sends letters of concern to the FCC about its policies on interconnection. When members of congress challenge the actions of the FCC this is considered an organizational challenge because they are seeking to protect the interests of their constituencies. |
| State rules established | A PUC clarifies or determines rules to be followed by carriers that provide services in that state. |
| Company refusal for interconnection | When, as reported by third parties to a government agency, a carrier does not facilitated interconnection with competing companies. When it is necessary for a government agency to intervene in a complaint about a denial for interconnection. CLEC complaints about failure to obtain interconnection under sec. 251. High rates are considered unwillingness to allow interconnection. Interconnection disputes among companies are considered unwillingness to provide interconnection. |
| Individual company review | When the FCC or a PUC reviews a company's application for interconnection related issues such as requests for certain networks or tariffs for interconnection to be exempted. When an issue arises that only pertains to the company under review. Sec. 271 approval is considered an individual company review. The review can include satellite companies. When companies complain about another company that obtains certain conditions of which they disapprove. |
| Federal rules yet to be determined | Companies complain about an issue that the FCC has not yet decided. |

## Appendix 2
### Universal service: operationalization of codes

| US funds received | When it is stated that a company received funds from the US fund. Also when the organization that provides the funds states the amount of funds available for support. Funds can also be made available to individuals. When there are announcements concerning the distribution of funds and potential concerns regarding eligible it items. |
|---|---|
| State rules to be determined | When state legislation is still being written or there is debate around an issue that a state has to settle. |
| Federal rules established | When the FCC sets up organizations to manage the USF. When there is a decision in the Senate or House about universal service. Speeches are not coded. Studies by consulting companies are not coded. |
| Not coded | Administrative information such as data requests from carriers is not coded. Journalist opinions are not coded. |